\documentclass[
    ,final            % use final for the camera ready runs
  ]
  {aipproc}

\layoutstyle{6x9}

\begin{document}

\title{Polarization puts a New Spin on Physics}

%http://www.aip.org/pacs/index.html

\author{John Ellis}{
  address={Theory Division, Physics Department, CERN, CH-1211 Geneva 23, Switzerland}
}

%\author{<author2>}{
% address={<common address for author2 and author3>}
%}

\begin{abstract}
Polarization and spin effects are useful for probing the Standard Model, in both the
electroweak sector and the strong sector, where the spin decomposition of the nucleon
is still a hot topic, with important new data on the net polarizations of the
gluon and the strange quarks. Spin phenomena are also useful in searches for
new physics, for example via measurements of the anomalous magnetic moment of the muon
and searches for electric dipole moments. The cross sections for the direct
detection of dark matter may also have an important spin-dependent component,
related to the spin decomposition of the nucleon, that could be an important diagnostic
tool. Polarization effects are also important diagnostic aids for high-energy
experiments at electron-proton, proton-proton and electron-positron colliders.
\end{abstract}

\maketitle

%%%%%%%%%%%%%%%%%%%%%%%%%%%%%%%%%%%%%%%%%%%%
%% MAINMATTER
%%%%%%%%%%%%%%%%%%%%%%%%%%%%%%%%%%%%%%%%%%%%
\begin{center}
CERN-PH-TH/2007-002
\end{center}

\section{Introduction}

Elementary particles have both external properties, 
namely the quantum numbers that determine
their couplings, and internal degrees of freedom associated with spin. These are often
linked: for example, the two helicity states of a fermion may have different electroweak
interactions. Polarization experiments and spin observables may therefore provide key 
insights into the properties of particles and their interactions. There are many examples
illustrating how such spin effects are helping to elucidate the Standard Model, and there
are justified hopes that they may help identify new physics beyond the Standard Model.
Indeed, the anomalous magnetic moment of the muon may already be providing the
first hints for new physics.

In this talk, I preview some of the hot topics for discussion at this conference, including
polarization phenomena in the electroweak interactions, the continuing puzzle of the
nucleon spin, searches for new physics via spin effects in low-energy experiments, and
the prospects for polarization experiments in high-energy collider experiments.
Spin and polarization effects are integral parts of the experimental toolkit for
disentangling new physics.

\section{Probing the Standard Model}

\subsection{Electroweak Physics}

Many of the most sensitive tests of the electroweak sector of the Standard Model
are being made with polarized beams, or using final-state spin effects~\cite{Erler}.
For example, one of the most accurate individual measurements at the $Z$ pole
has been the $A_{LR}$ asymmetry measured at SLAC using polarized beams,
and one of the most precise LEP determinations of $\sin^2 \theta_W$ was made
using final-state $\tau$ polarization. 

\begin{figure}
 \includegraphics[width=0.38\textwidth,angle=90]{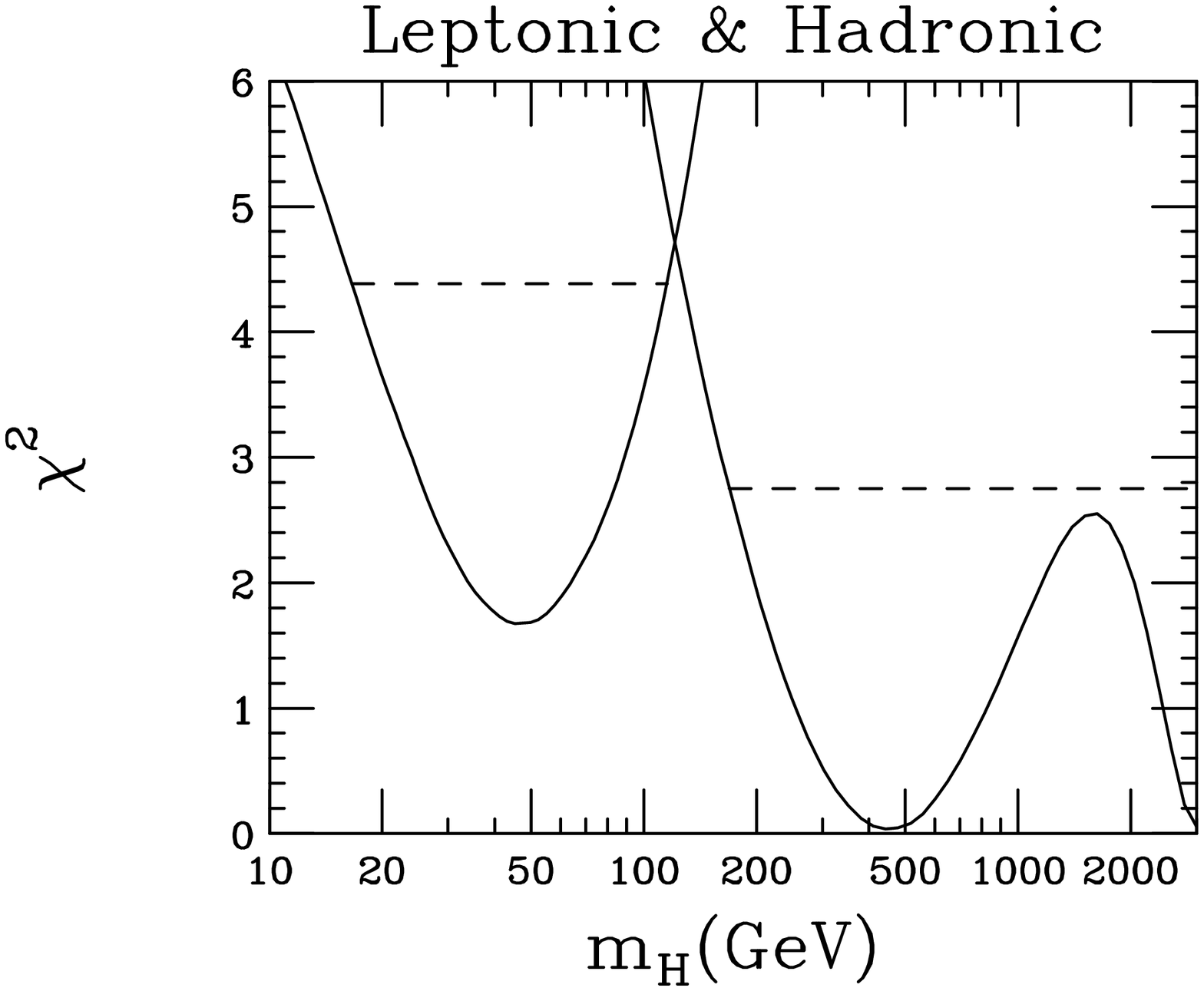}
  \hfill
 \includegraphics[width=0.52\textwidth]{s2w_2006.eps}
  \caption{
     As seen in the left panel, measurements of $\sin^2 \theta_W$ using leptonic
     asymmetries prefer a low value of $m_H$, whereas hadronic measurements
     prefer a higher value~\protect\cite{Chanowitz}. 
     The right panel shows present and prospective probes of
     the energy dependence of $\sin^2 \theta_W$, including Moeller scattering,
     parity violation in atomic physics, and asymmetry 
     measurements at high energy~\protect\cite{Erler}.}
\label{fig:CE}
\end{figure}

The name of the electroweak game now is to use the precision
measurements to estimate the mass of the Higgs boson. 
As seen in Fig.~\ref{fig:CE}, the leptonic asymmetries tend to prefer a
relatively low value of the Higgs mass, but the agreement between these and the
hadronic determinations of $\sin^2 \theta_W$ is not perfect, and the latter prefer a
heavier Higgs mass~\cite{Chanowitz}. The conventional attitude towards this discrepancy is to
regard it as a statistical fluctuation, and to ignore it in performing a global fit to the
precision electroweak data, which leads to a relatively low Higgs mass:
$m_H = 85^{+39}_{-28}$~GeV~\cite{EWWG}. 
However, it is also possible that the discrepancy is
real, and betokens new physics, in which case this Standard Model estimate would
be invalid~\cite{Chanowitz}. A direct resolution of this issue is unlikely in the near future, 
since a high-precision return to the $Z$ peak is not foreseen unless and until the
Giga-$Z$ option (with polarized electrons and positrons)
is exercised at the ILC~\cite{GigaZ}. Perhaps the issue will be resolved by the
detection of the Higgs boson at the LHC, and the comparison of its mass with the
prediction of the global electroweak fit and/or subsets of the electroweak data?

For the time being, the action in
precision electroweak tests is in measurements at both low and high energies,
aiming to check the specific energy dependence predicted by radiative corrections
in the Standard Model~\cite{Erler}. As also seen in Fig.~\ref{fig:CE},
there are puzzles below the $Z$ pole, such as the anomalous value of $\sin^2 \theta_W$
extracted from deep-inelastic $\nu$-N scattering~\cite{NuTeV}. This might also indicate some
problem with the Standard Model, although the interpretation of this 
experiment is vulnerable to hadronic uncertainties. However, this discrepancy
certainly increases the interest in other
low-energy experiments with different systematic uncertainties. 

One particularly clean
measurement is that of parity violation in Moeller scattering~\cite{Moeller}, which is theoretically very clean.
The new SLAC measurement agrees with the Standard Model to within about one standard
deviation, and a new measurement with greater accuracy would be particularly interesting.
Also interesting are measurements of measurements of atomic parity violation,
which are currently also within about one standard deviation of the Standard Model
prediction, and show potential for
improved accuracy in the future. Above the $Z$ pole, we can also expect future 
measurements to improve on the current determination of $\sin^2 \theta_W$ using the
forward-backward asymmetry at $Q > 1000$~GeV.

The few examples given above illustrate the importance of spin and polarization
experiments in the analysis of elecotroweak physics.

\subsection{The Nucleon Spin Puzzle}

\begin{figure}
 \includegraphics[width=0.34\textwidth,angle=90]{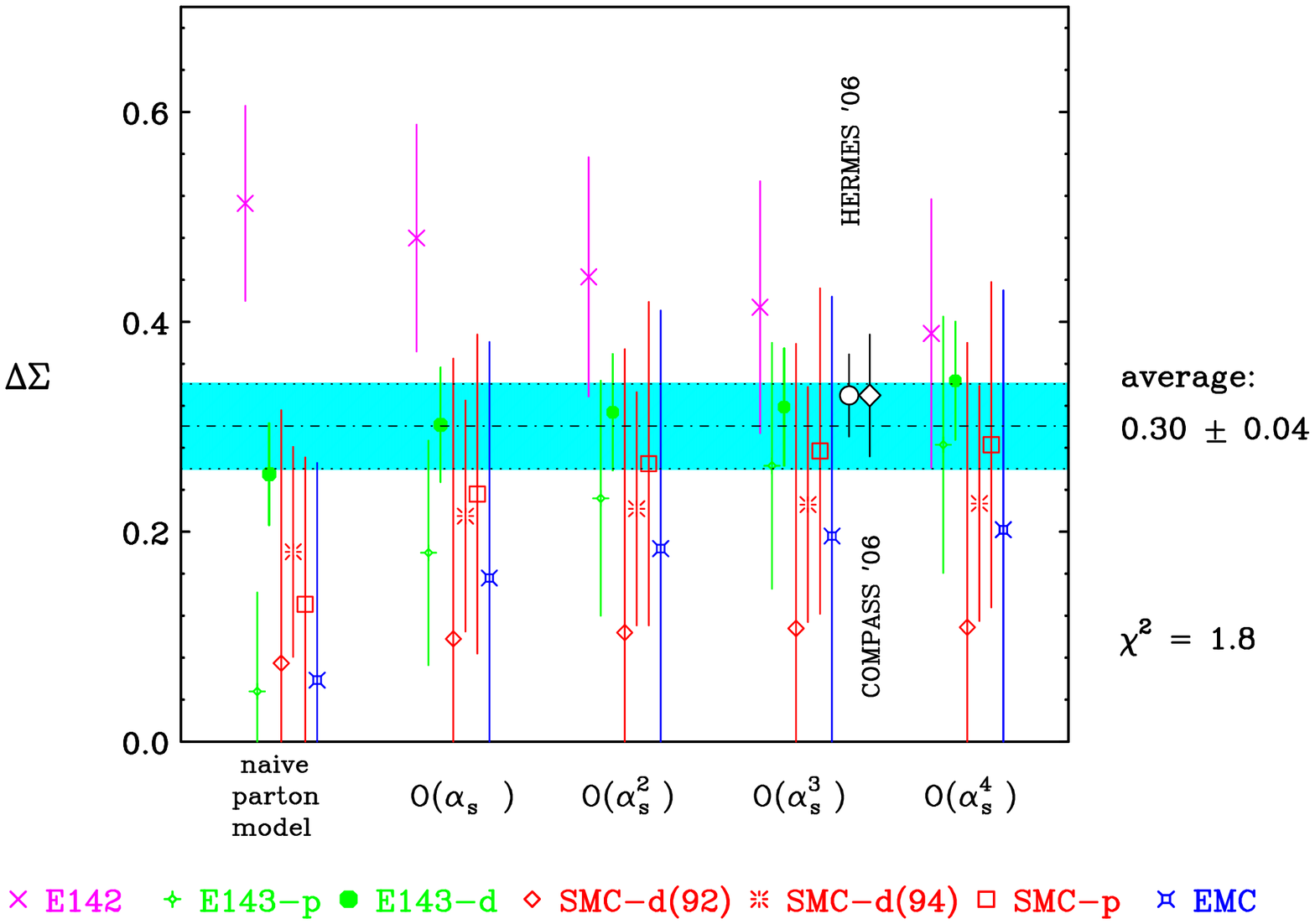}
  \hfill
 \includegraphics[width=0.56\textwidth]{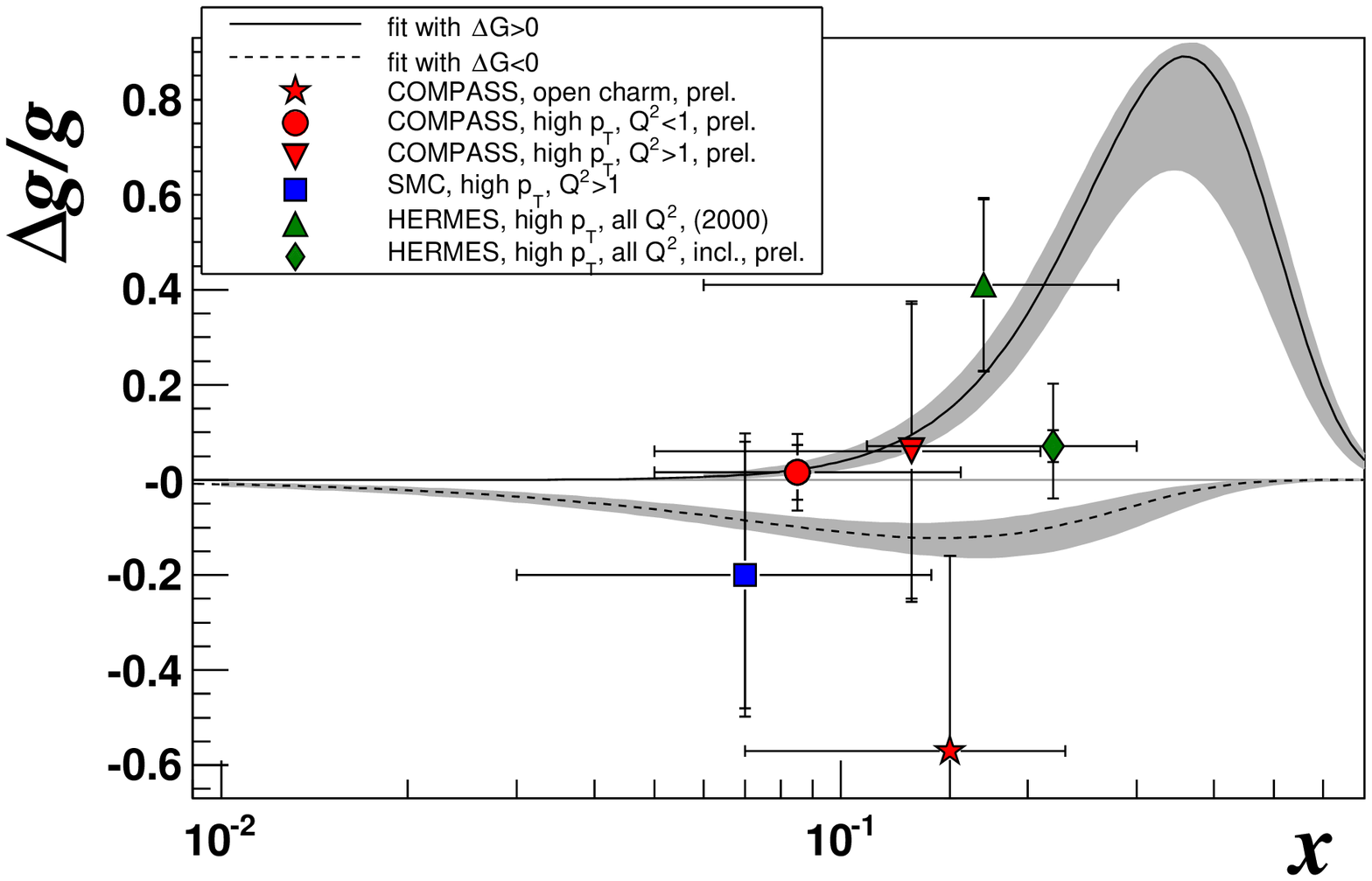}
  \caption{\label{fig:EK}
     As seen in the left panel, measurements of the total quark contribution to the
     nucleon are remarkably consistent and stable, once higher-order perturbative corrections are
     taken into account~\protect\cite{EK}. 
     The right panel shows current measurements of $\Delta G$ by the
     SMC, HERMES and COMPASS Collaborations. The latter, in particular, indicate that
     $\Delta G$ is unlikely to be very large~\protect\cite{Mallot}.}
\end{figure}

There are several approaches to this problem, in particular via measurements of
scaling violations, via jet and charm production
asymmetries in polarized deep-inelastic scattering, and via asymmetries in polarized
proton-proton scattering. The COMPASS collaboration has recently presented new data on the 
deuteron polarized structure function $g_1^d(x)$~\cite{COMPASS}. 
These provide significant improvements at
small $x$, in particular, and have interesting sensitivity to the net polarizations
of the strange quarks, $\Delta s$, and the gluons, $\Delta G$.
The HERMES collaboration has also presented new data on 
$g_1^p(x)$ and $g_1^d(x)$~\cite{HERMES}, 
which provide significant improvements at large $x$. These two data sets tell strikingly
similar stories about the singlet axial-current matrix element $a_0$, which is related
to the total net quark contribution $\Delta \Sigma$ to the nucleon spin:
\begin{eqnarray}
a_0 (Q^2 = 3 {\rm GeV}^2) & = & 0.35~ \pm 0.03~ ({\rm stat.}) \pm 0.05~ ({\rm syst.}) {\rm [COMPASS]}, \\
a_0 (Q^2 = 5 {\rm GeV}^2) & = & 0.330 \pm 0.011 ({\rm th.}) \pm 0.025 ({\rm exp.}) 
\pm  0.028({\rm ev.}) {\rm [HERMES]}.
\end{eqnarray}
The value of $a_0$ has remained quite stable and consistent over the past decade, ever since 
SLAC and CERN data on deep-inelastic electron and muon scattering on proton and deuteron
targets were first compared including perturbative ${\cal O}(\alpha_s^3)$ contributions to the
integrals over $g_1^p(x)$ and $g_1^d(x)$~\cite{EK}, as seen in Fig.~\ref{fig:EK}.

\begin{figure}[t]
\includegraphics[width=0.425\textwidth]{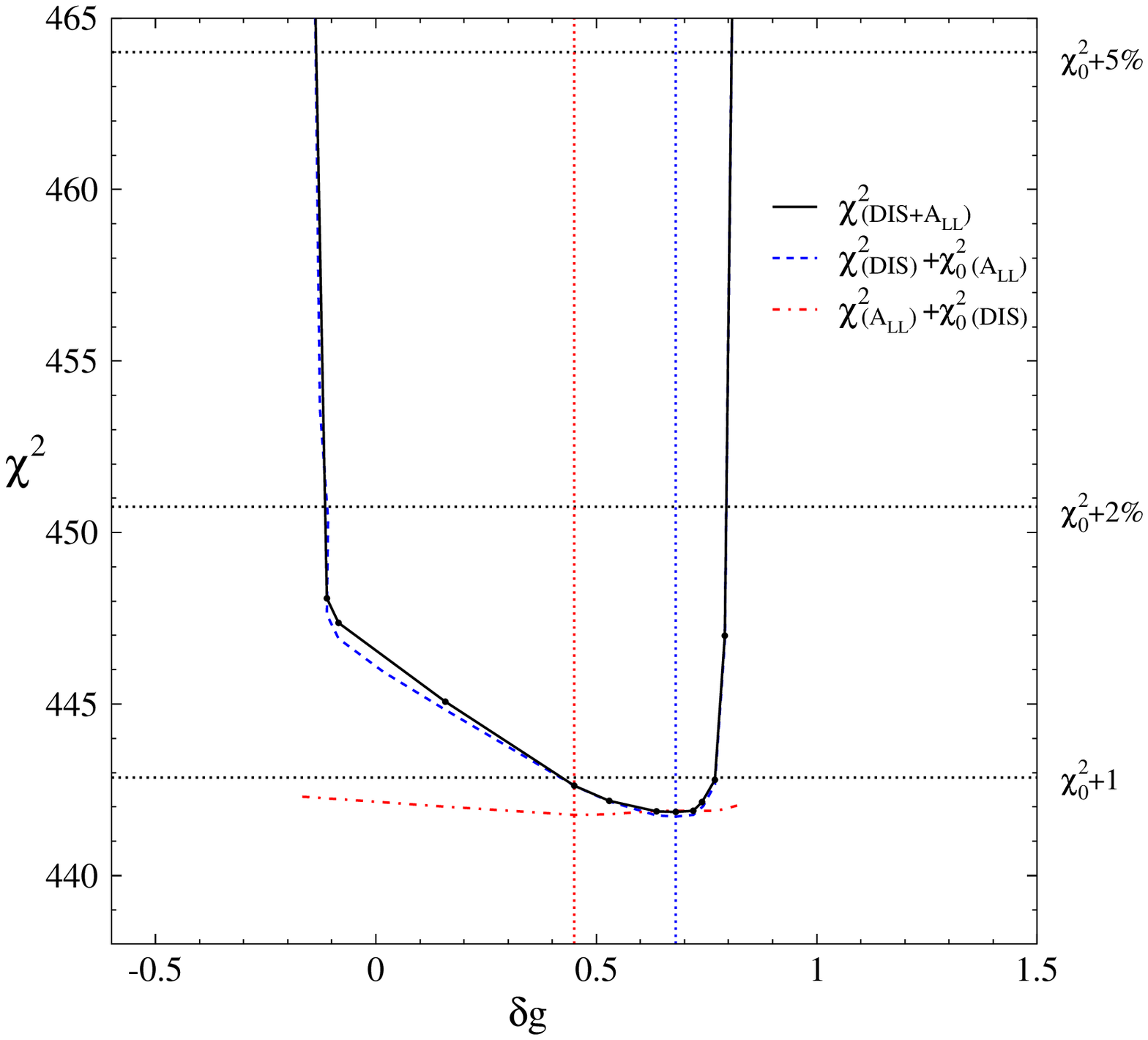}
\includegraphics[width=0.475\textwidth]{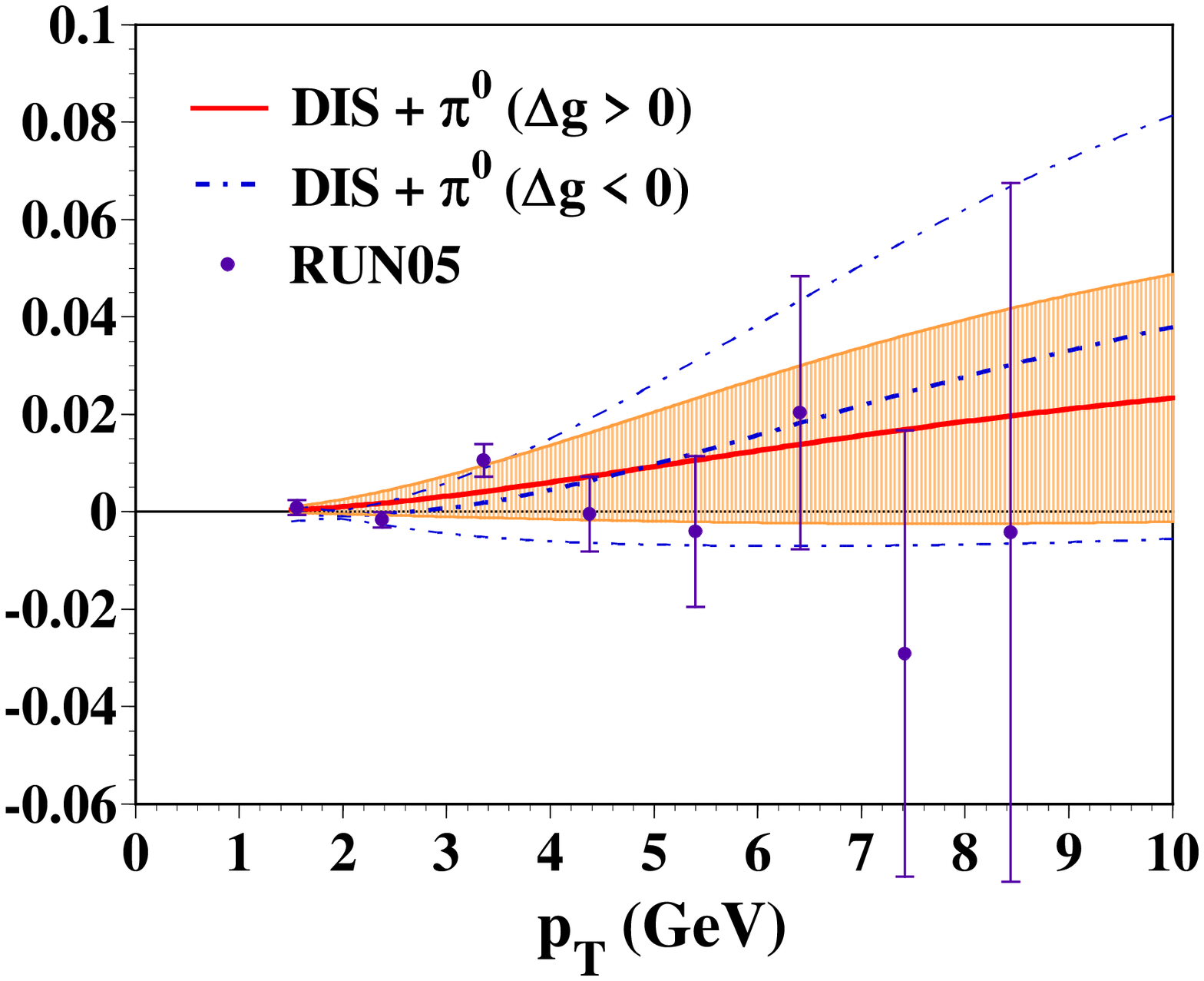}
  \caption{\label{fig:Gabriela}
     Left panel: Contributions to the $\chi^2$ function for
     $\Delta G$ due to positivity, deep-inelastic inclusive data (blue dashed line),
     earlier RHIC data (red dash-dotted line) and total (black solid line)~\protect\cite{Gabriela}.
     Right panel: Indications on $\Delta G$ from more 
     recent PHENIX data from RHIC~\protect\cite{AAC}.}
\end{figure}

Correspondingly, the two experiments also tell very similar stories for the net
contribution to the nucleon spin of the strange quarks and antiquarks:
\begin{eqnarray}
\Delta s & = & -0.08~ \pm -0.01 ({\rm stat.}) \pm 0.02 ({\rm syst.}) {\rm [COMPASS]}, \\
\Delta s & = & -0.085 \pm 0.013 ({\rm th.}) \pm 0.008 ({\rm exp.}) 
\pm  0.009({\rm ev.}) {\rm [HERMES]}.
\label{deltas}
\end{eqnarray}
According to these experiments, $\Delta s$ is significantly negative.
The measurements of scaling violations in structure functions
also indicate that $\Delta G$ is probably not large, as seen in Fig.~\ref{fig:Gabriela},
and may favour small positive values~\cite{Gabriela}.

More direct information on the possible magnitude of $\Delta G$ has been emerging from a series of
measurements by the SMC, HERMES, COMPASS and RHIC Collaborations. 
As seen in Fig.~\ref{fig:EK}, the early SMC and
COMPASS measurements of jet asymmetries and open charm production had quite large errors, 
and indicated only that $\Delta G$ was unlikely to be very large and positive:
$\Delta G \ll G$~\cite{Mallot}. The most accurate
information on $\Delta G$ is now that provided by the new COMPASS measurement of the jet
asymmetry at low $Q^2$, based on data taken between 2002 and 2004. It indicates that $\Delta G(x)$
must be close to zero in a range of $x \sim 0.085$ for $\mu^2 \sim 3$~GeV$^2$~\cite{COMPASS2}. 
A recent
update of the earlier HERMES measurement also indicates a small value of $\Delta G(x)$ in a range of 
$x \sim 0.22$ for $\mu^2 \sim 1.35$~GeV$^2$~\cite{HERMES2}:
\begin{eqnarray}
\Delta G & = & 0.016 \pm 0.058 ({\rm stat.}) \pm 0.055 ({\rm syst.}) {\rm [COMPASS]}, \\
\Delta G & = & 0.071 \pm 0.034 ({\rm stat.}) ^{+ 0.105}_{- 0.127} ({\rm syst.}) {\rm [HERMES]}.
\label{deltag}
\end{eqnarray}
In view of the results in Fig.~\ref{fig:EK} and the positivity constraints shown in
Fig.~\ref{fig:Gabriela}, it seems inescapable that $\Delta G$ cannot be 
large enough for gluon renormalization effects~\cite{glueren}
to be able to make any significant contribution to $a_0$~\cite{Mallot}, even forgetting about the
renormalization-scheme ambiguity~\cite{JM}. COMPASS obtained significantly more data in 2006, in
particular on open charm production, and one may hope that these will cast more light on the
size of $\Delta G$. The next step will be to establish whether $\Delta G$ makes any significant
contribution to the nucleon spin, or whether its contribution is negligible, i.e., whether
$\Delta G$ is closer to 1/2 or to zero.

In this respect, the main future competition for COMPASS will come from the RHIC 
experiments~\cite{STAR,AAC}. Currently.
these also indicate that $\Delta G \ll G$, but do not yet provide qualitative additional information, 
in particular because their measurements of $A_{LL}$ depend quadratically on $\Delta G$.
However, the RHIC data taken in 2006 should provide significant extra information.

My list of the main issues for the future includes the following.

$\bullet$ Obtain direct information on the magnitude and sign of $\Delta s$, e.g., from
asymmetries in strange particle production. The pioneering data from HERMES~\cite{HERMs}
are not at
large enough $Q^2$ and $W^2$ to be sure that asymptotic factorizing models for
fragmentation can be applied easily~\cite{Aram}.

$\bullet$ Determine whether $\Delta G$ provides a large fraction of the nucleon spin, e.g., via
production asymmetries in deep-inelastic scattering or at RHIC.

$\bullet$ Determine the magnitude of the orbital angular momentum component, e.g., by
measurements of deeply-virtual Compton scattering.

$\bullet$ Forge better theoretical connections with other spin-dependent observables, such
as particle production in low-energy proton-antiproton annihilation and $\Lambda$
production in deep-inelastic scattering. Data on the polarization of $\Lambda$ baryons
produced in deep-inelastic scattering are suggestive, but subject to significant theoretical
uncertainties.

\section{Looking for New Physics beyond the Standard Model}

The standard list of questions in particle physics beyond the Standard Model
includes the following.

$\bullet$ What is the origin of particle masses? Are they due to a Higgs boson, and
is this accompanied by other new physics such as supersymmetry? These questions 
are likely to be answered at energies below $\sim 1$~TeV.

$\bullet$ Why are there so many different types of matter particles? How is this issue related
to the small CP-violating matter-antimatter difference that has been seen in the laboratory, and can
the answer to this question be related to the cosmological dominance of matter over antimatter?

$\bullet$ Are the fundamental forces unified? If so, unification may occur only at some
very high energy $\sim 10^{16}$~GeV. Direct probes of unification include neutrino physics
and the search for baryon decay, but indirect probes are also possible at colliders
via measurements of particle masses and couplings.

$\bullet$ How can one formulate a quantum theory of gravity? Is it based on (super)string theory,
in which case are there large extra space-time dimensions?

As discussed below, spin and polarization phenomena may play important roles in many of the
searches for new physics. For example, spin correlations may be crucial for distingushing
between supersymmetry and some scenarios with large extra dimensions. The anomalous
magnetic dipole moment of the muon may already be providing us with a hint of new physics,
that could be comaptible with supersymmetry. Electric dipole moments are key probes of
CP violation. Spin-dependent interactions may also be provide valuable tools in the search for
astrophysical dark matter.

\subsection{An example of possible new physics}

Supersymmetry is the last undiscovered particle symmetry. The first reason why it appeared
interesting was because it could link particles with spins differing by half a unit,
namely fermions and bosons, and hence unify matter and 
force particles. With sufficiently many supersymmetries, one could relate particles of all
different spins, including Higgs-like particles with spin 0, matter particles with spin 1/2,
gauge particles with spin 1, the gravitino with spin 3/2, and the graviton with spin 2.
This very elegant motivation gave, however, no indication of the possible mass scale of
the supersymmetric partners of the particles of the Standard Model. The first indication
that their masses might be around a TeV was provided by the observation 
that in this case they could help fix the
electroweak masses, by controlling the loop corrections that would otherwise destroy
the hierarchy of fundamental mass scales~\cite{hierarchy}. 
Later motivations that supersymmetry might appear
at the TeV scale were provided by its potential help in achieving grand unification~\cite{GUT}, 
and the
possibility that the lightest supersymmetric particle (LSP) might provide the astrophysical
dark matter~\cite{EHNOS}. Since supersymmetry involves spin in an essential way, it should 
come as no
surprise that many promising ways to probe the theory involve observables based on
spin and/or polarization.

\subsection{The muon anomalous magnetic moment}

The anomalous magnetic moment of the muon, $a_\mu = (g_\mu - 2)/2$ may already be
providing the first accelerator evidence for new physics. The measurement by the BNL
g-2 Collaboration~\cite{g-2} disagrees significantly with the Standard Model if $e^+ e^-$
annihilation data are used to calculate the Standard Model contribution, although there is no
significant discrepancy if this is calculated using $\tau$-decay data~\cite{Davier}. The jury has yet to
deliver its verdict, but the weight of evidence is accumulating on the side of the $e^+ e^-$
data and hence in favour of a hint for new physics. There are several new sets of low-energy 
$e^+ e^-$ data, which have a good level of consistency. In contrast, new $\tau$-decay data
from the BELLE Collaboration seem significantly different from the previous ALEPH and CLEO
data, and agree better with the $e^+ e^-$ data. If the $e^+ e^-$ estimate of the hadronic
contribution to the Standard Model calculation is accepted, the theoretical error is just a
few per cent, and is comparable to the experimental error~\cite{ICHEP}:
\begin{eqnarray}
a_\mu ({\rm theory}) & = & (11659180.5 \pm 5.6) \times 10^{-10}, \\
a_\mu ({\rm experiment}) & = & (11659208.0 \pm 6.3) \times 10^{-10}, \\
\Delta a_\mu & = & (27.5 \pm 8.4)  \times 10^{-10},
\label{amu}
\end{eqnarray}
yielding a discrepancy at the 3.3-$\sigma$ level, that could be due to supersymmetry~\cite{susyg-2},
for example.

\subsection{Electric dipole moments}

Aspects of supersymmetry can also be probed by searches for electric dipole moments,
whose existence would be direct evidence for CP and T violation. They are expected to be
unobservably small in the Standard Model, which makes them ideal for searches for new physics.
In particular, they are sensitive to the many CP-violating phases in supersymmetric models, some of
which could be responsible for the dominance of matter over antimatter in the Universe.
Interesting upper limits on these CP-violating phases are already provided by the present upper 
limits on electric dipole moments:
\begin{eqnarray}
|d_{Tl}| & < & 9 \times 10^{-25}~e{\rm cm}, \\
|d_{Hg}| & < & 2 \times 10^{-28}~e{\rm cm}, \\
|d_{n}| & < & 6 \times 10^{-26}~e{\rm cm},
\label{edmnow}
\end{eqnarray}
and even more interesting would be the prospective sensitivities:
\begin{eqnarray}
|d_{e}| & < & 3 \times 10^{-29}~e{\rm cm}, \\
|d_{D}| & < & 2 \times 10^{-27}~e{\rm cm}, \\
|d_{n}| & < & 1 \times 10^{-27}~e{\rm cm},
\label{edmsoon}
\end{eqnarray}
obtainable in upcoming experiments.

The relevant CP-violating parameters in an effective low-energy Lagrangian include the
CP-violating strong-interaction phase $\theta$:
\begin{equation}
{\cal L}_{eff} \; \ni \; \frac{g_s^2}{32 \pi^2} \theta G_{\mu \nu}^a {\widetilde G^{\mu \nu, a}},
\label{theta}
\end{equation}
conventional and colour electric dipole moments of elementary fermions:
\begin{equation}
{\cal L}_{eff} \; \ni \; - \frac{1}{2} \Sigma_{i=e,u,d,s} d_i {\bar \psi_i} (F \sigma) \gamma_5 \psi_i 
- \frac{1}{2} \Sigma_{i=u,d,s} {\tilde d_i} {\bar \psi_i (F \sigma) \gamma_5 \psi_i } ,
\label{edms}
\end{equation}
a three-gluon operator:
\begin{equation}
{\cal L}_{eff} \; \ni \; \frac{1}{3} f^{abc} w G_{\mu \nu}^a {\widetilde G^{\nu \beta, b}} G_\beta^{\mu,c},
\label{threeg}
\end{equation}
and four-fermion operators:
\begin{equation}
{\cal L}_{eff} \; \ni \; \Sigma_{i,j} C_{ij} ({\bar \psi_i} \psi_i) ({\bar \psi_j} i \gamma_5 \psi_j) + \dots
\label{fourf}
\end{equation}
Each of the CP-violating parameters $\theta, d_i, {\tilde d_i}, w, C_{ij}$ may receive
contributions from the underlying CP-violating parameters of some extension of the
Standard Model, such as supersymmetry. For example, the electron electric dipole
moment $d_e$ is given at the one-loop level by~\cite{Pospelov}:
\begin{equation}
d_e = \frac{e m_e}{16 \pi^2 M_{SUSY}^2} \left[ \left(\frac{5g_2^2}{24} + \frac{g_1^2}{24}\right)
\tan \beta \sin\theta_\mu + \frac{g_1^2}{12}\sin \theta_A \right],
\label{eedm}
\end{equation}
where $\theta_\mu \equiv {\rm Arg}(\mu M_2m^{2*}_{12})$ and $\theta_A \equiv {\rm Arg}(M_1^*A_e)$
are two CP-violating relative phases between supersymmetric model parameters.

\begin{figure}
\centerline{\includegraphics[width=0.45\textwidth]{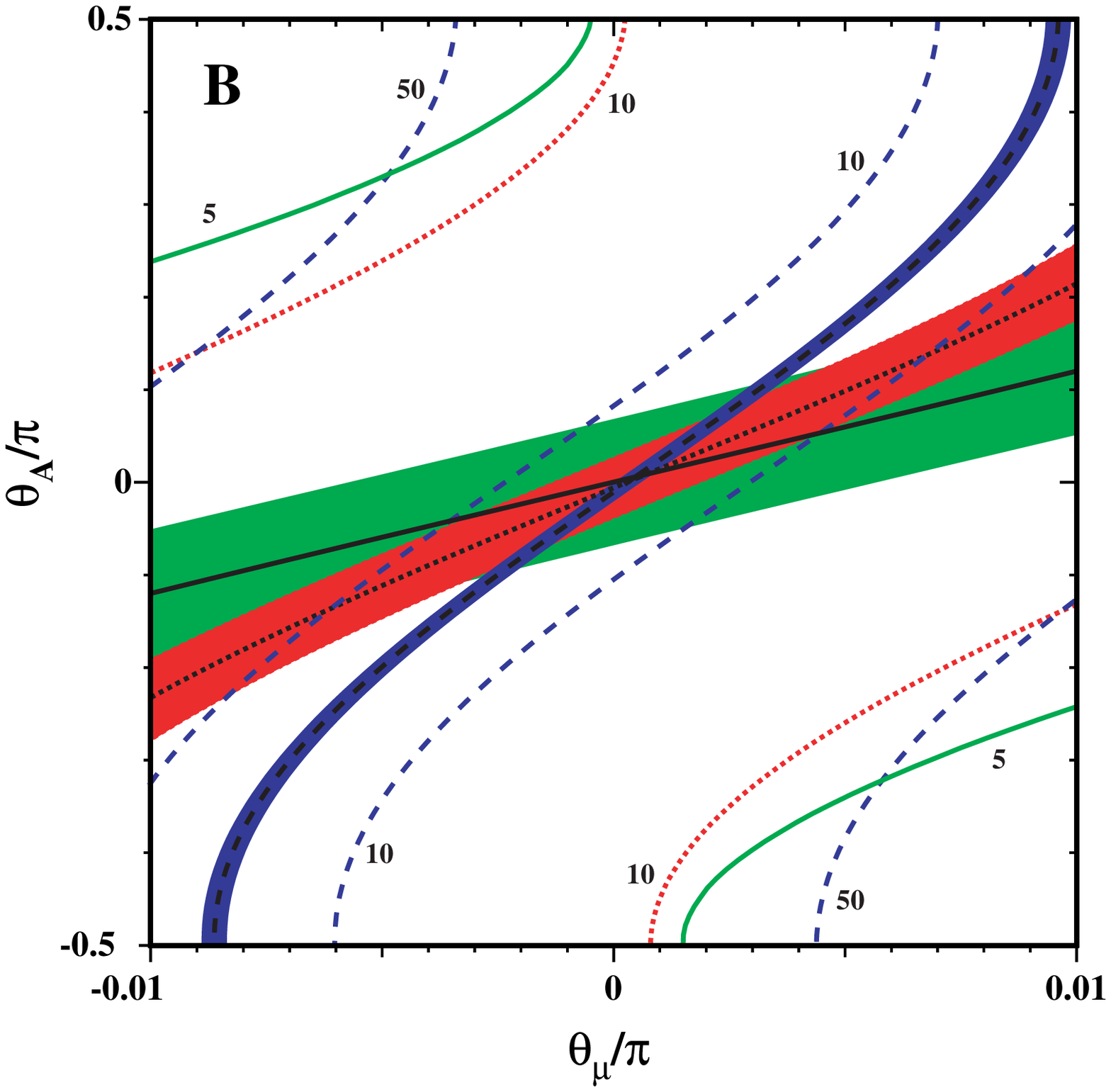}
\includegraphics[width=0.45\textwidth]{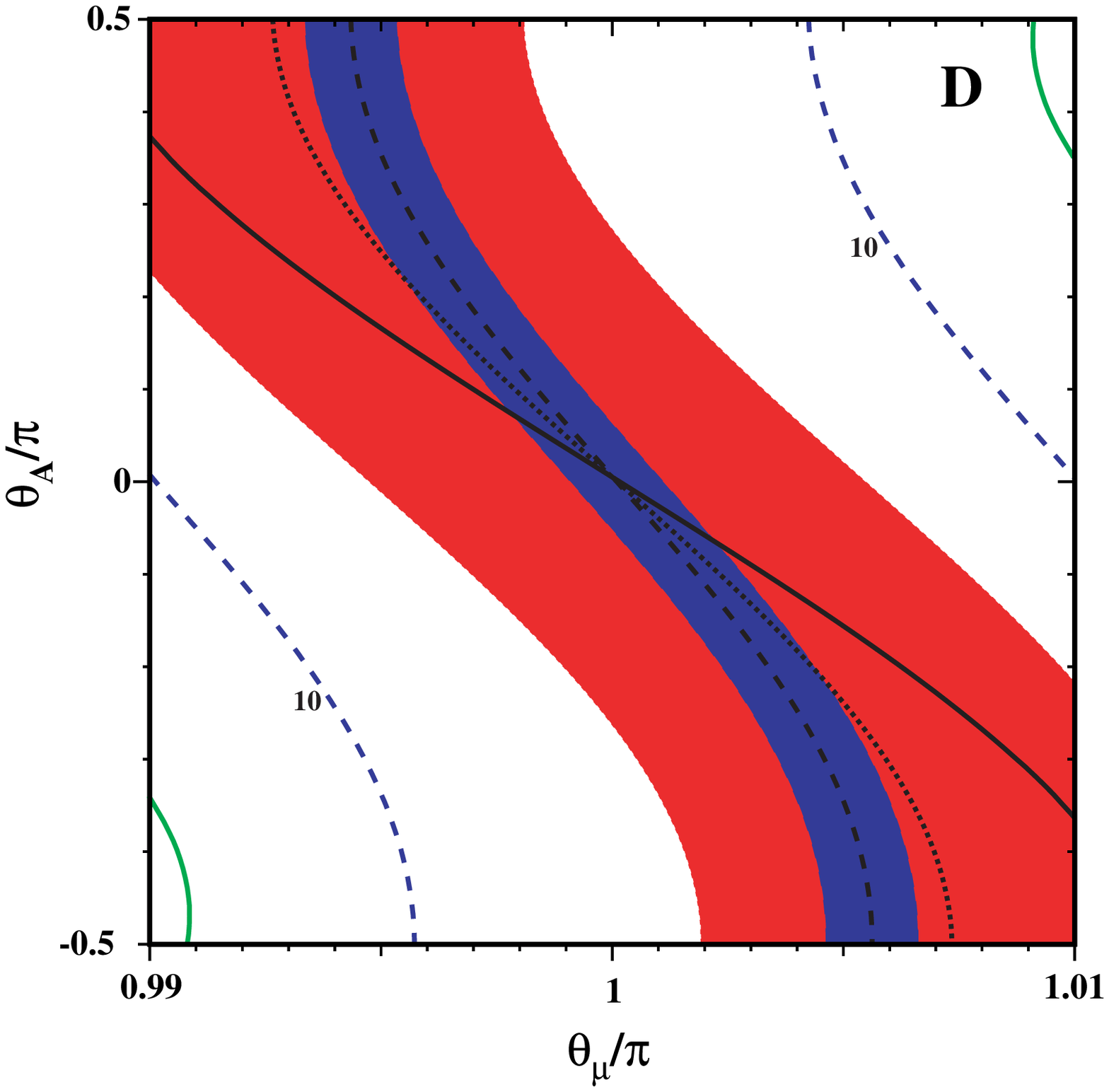}}
 \caption{
 Constraints on the CP-violating supersymmetric phases $\theta_\mu, \theta_A$ due to
 upper limits on the electric dipole moments of Thallium (blue dashed), the neutron (red dotted), and
 Mercury (green solid)~\protect\cite{Pospelov} for the supersymmetric 
 benchmark points B and D~\protect\cite{Bench}.  
 }
\label{fig:edms} 
\end{figure}

In the case of Thallium, the contributions from the electric dipole moment of the electron and
from four-fermion interactions are~\cite{Pospelov}
\begin{equation}
d_{Tl} \; \ni \; - 585 d_e - e.43 {\rm GeV}.C_S,
\label{Thallium}
\end{equation}
where $C_S$ contains a contribution from possible four-fermion interactions connecting
the electron to $d$ and heavier quarks: $C_S \ni C_{de} (29 {\rm MeV}/m_d) + \dots$. In the
case of the neutron, the contributions of the electric dipole moments of the quarks are
\begin{equation}
d_n \; \ni \; (1.4 \pm 0.6)(d_d - 0.25 d_u) + (1.1 \pm 0.5)({\tilde d}_d + 0.5 {\tilde d}_u) e .
\label{neutron}
\end{equation}
Fig.~\ref{fig:edms} demonstrates~\cite{Pospelov} the combined impacts of present experimental upper
limits on various electric dipole moments as functions of the CP-violating
phases $\theta_A$ and $\theta_\mu$ for representative benchmark values of other
supersymmetric model parameters~\cite{Bench}. In the case of model B, for example,
one has $|\theta_A/\pi| < 0.08, |\theta_\mu/\pi| < 0.002$, and for model D
there is no upper limit on $\theta_A$, whereas $|\theta_\mu/\pi| < 0.07$.
These results illustrate that $\theta_A$ may well be ${\cal O}(1)$,
whereas $\theta_\mu$ is likely to be smaller than about 0.2. Unfortunately, there is
no real theoretical guidance on the possible magnitudes of these
supersymmetric phases, which are not related in any obvious way to the
Kobayashi-Maskawa phase. However, the pressure on models will certainly increase as
the experimental sensitivities to electric dipole moments are improved.

An interesting new idea is to measure the deuteron electric dipole moment
to an accuracy $\sim 10^{-29}$~e.cm using forced oscillations of particle velocities in
resonance with spin precession in a 1.5 GeV storage ring~\cite{Orlov}. Since one
expects that $d_D \; = \; d_n + d_p + \dots$, this experiment would provide a very
interesting improvement in sensitivity over the present and planned neutron electric dipole moment
experiments.
  
\subsection{Searching for supersymmetric dark matter}

As already mentioned, in many supersymmetric models, the LSP $\chi$ is stable, 
and a suitable candidate for astrophysical dark matter~\cite{EHNOS}. Several strategies to
search for LSP dark matter have been proposed, including looking for the products 
of $\chi \chi$ annihilations in the galactic halo, such as antiprotons and positrons, or for
energetic photons due to annihilations in the galactic centre. Other strategies
include looking for energetic neutrinos produced by annihilations inside the core
of the Sun or Earth. The rate for solar annihilations is generally controlled by the
capture rate, which is largely due to $\chi$ energy loss during scattering on protons inside the Sun:
$\chi + p \to \chi + p$. This scattering is in turn dominated by spin-dependent interactions that 
are sensitive to axial-current matrix elements related to the magnitude
of $\Delta s$~\cite{EFR}. This quantity may also be important for the direct search for dark matter
scattering on nuclei in the laboratory: $\chi + A \to \chi + A$.

The experimental upper limits on spin-independent scattering of dark matter particles
are beginning to eat into the parameter spaces of some supersymmetric models, as seen
in the left panel of Fig.~\ref{fig:CDM}~\cite{EOSS}. The limits on 
spin-dependent dark matter scattering currently lie far above the predictions in 
favoured supersymmetric models, as seen in the right panel of Fig.~\ref{fig:CDM}~\cite{EOSS},
but there are proposals to improve the experimental
sensitivity significantly. If ever a dark matter candidate is found, a measurement of spin-dependent
could be a key tool for diagnosing the underlying supersymmetric model.

\section{High-energy colliders}

\subsection{Electron-proton collisions}

HERA has recently started providing the first measurements of polarization asymmetries in
deep-inelastic high-energy $e^\pm$-proton collisions~\cite{ICHEP}, and will continue to take these data
until mid-2007. These data remove overall sign ambiguities in the vector and axial
couplings of the $u$ and $d$ quarks to the $Z$ boson, confirming that they have the
signs predicted in the Standard Model. This ambiguity has also been lifted by previous
unpolarized HERA data, but the determinations using polarized data are already much
more precise, particularly in pinning down the $u$ couplings to the $Z$ boson.

{\begin{figure}[t]
\centerline{\includegraphics[width=0.45\textwidth]{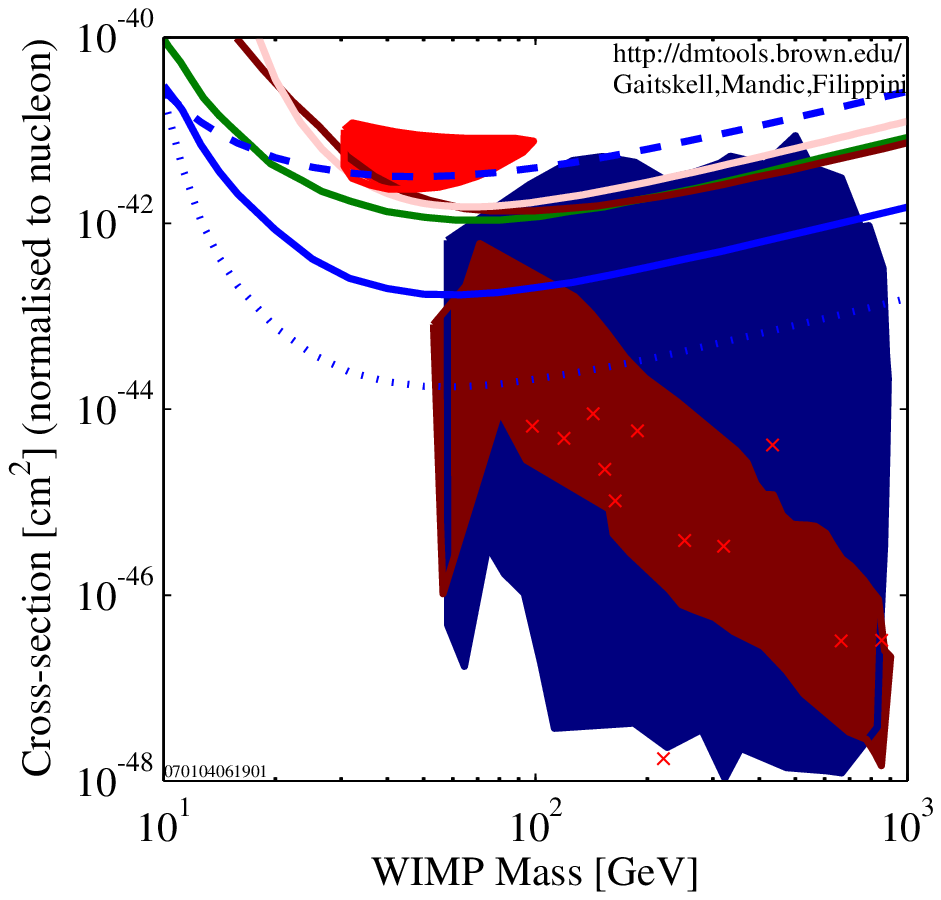}
\includegraphics[width=0.45\textwidth]{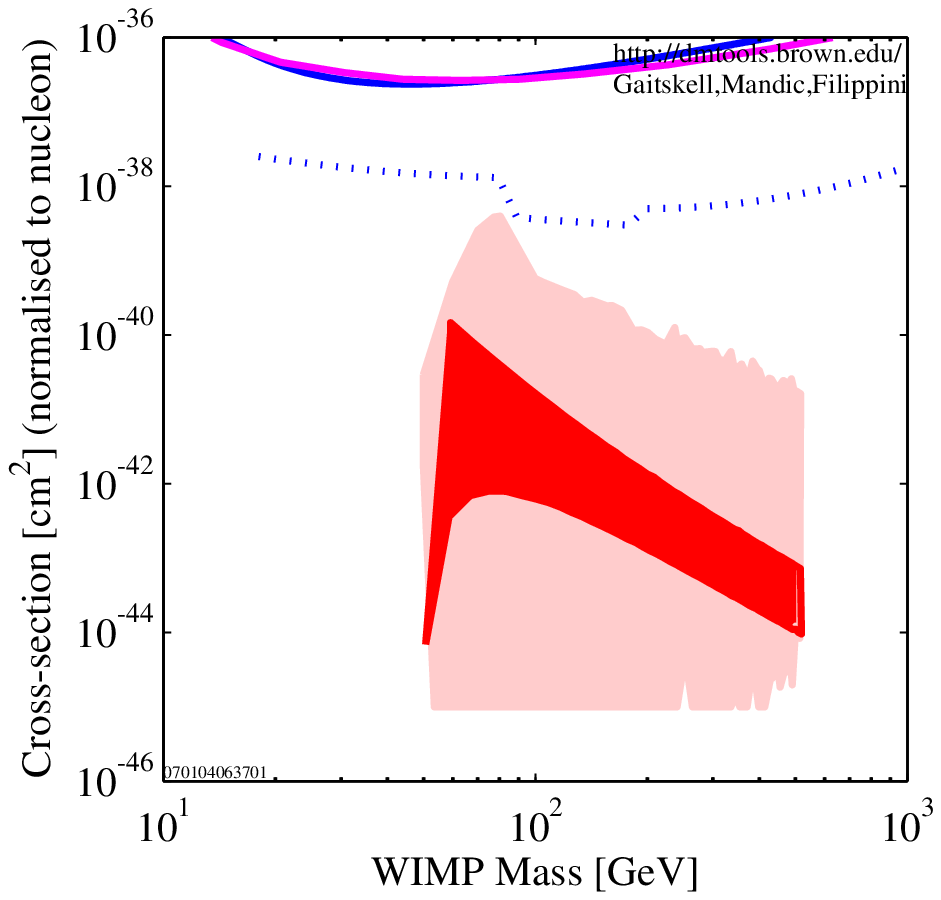}}
\end{figure}
\begin{figure}
\centerline{\includegraphics[width=0.45\textwidth]{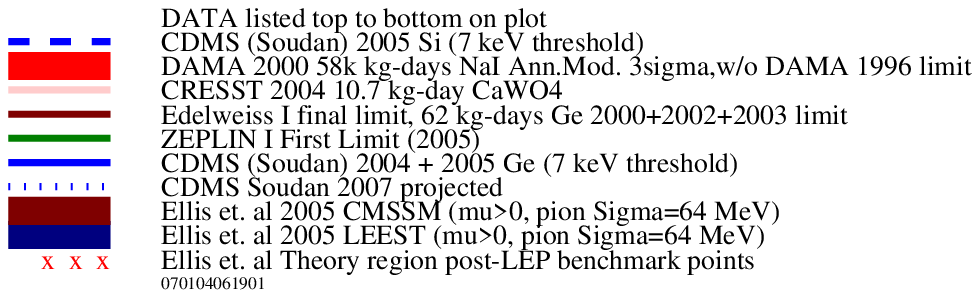}
\includegraphics[width=0.45\textwidth]{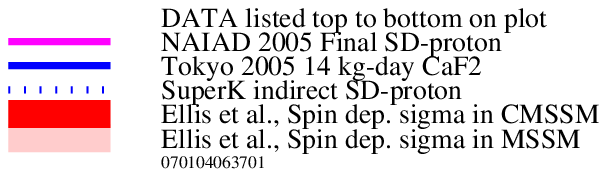}}
 \caption{
 Upper limits on spin-independent LSP scattering (left) and spin-dependent LSP-proton scattering
 (right), compared with theoretical estimates~\protect\cite{EOSS}. The plots are from~\protect\cite{GMF}.
 }
\label{fig:CDM} 
\end{figure}}

\subsection{Hadron-hadron collisions}

Although present and planned hadron colliders will not have polarized beams, there
is interesting information to be obtained from measurements of final-state
spins and their correlations. One example is provided by measurement of the
$W$ polarization in top quark decay. There are two observables, $f_{0,+}$, and the
Standard Model predicts that that only $f_0$ should be non-zero. Data from the CDF
and D0 Collaborations are indeed compatible with the Standard Model value for $f_0$
and $f_+ = 0$, providing a valuable check on the weak interactions of the $t$ quark~\cite{tWdecay}.

The LHC, now nearing completion at CERN, will collide proton beams of 7~TeV each
with a design luminosity of $10^{34}$~cm$^{-2}$s$^{-1}$. All of the magnets have now
been delivered to CERN, and most of them have been installed in the LHC tunnel. The
interconnection of these and other machine components is proceeding apace, and the
machine is expected to be closed by the end of August 2007, with first collisions
planned for November 2007. It will not be possible before then to commission the entire
machine for operation at the design energy, so the 2007 running will be at the injection
energy of 900~GeV in the centre of mass. The commissioning will be completed in the
first part of 2008, so that full-energy running can start by the middle of that year.

\begin{figure}
\centerline{\includegraphics[width=6.2cm]{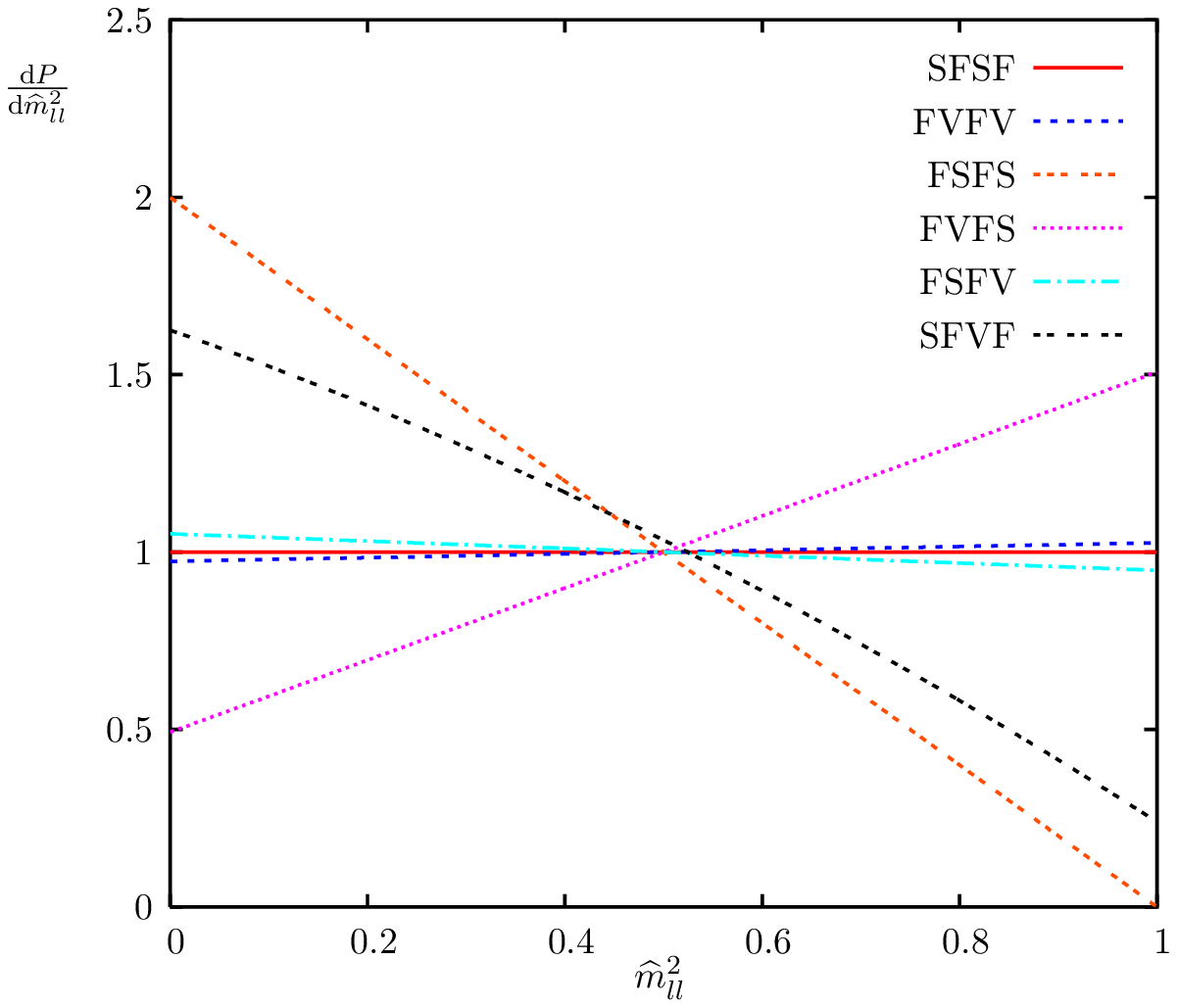}
\includegraphics[width=7.8cm]{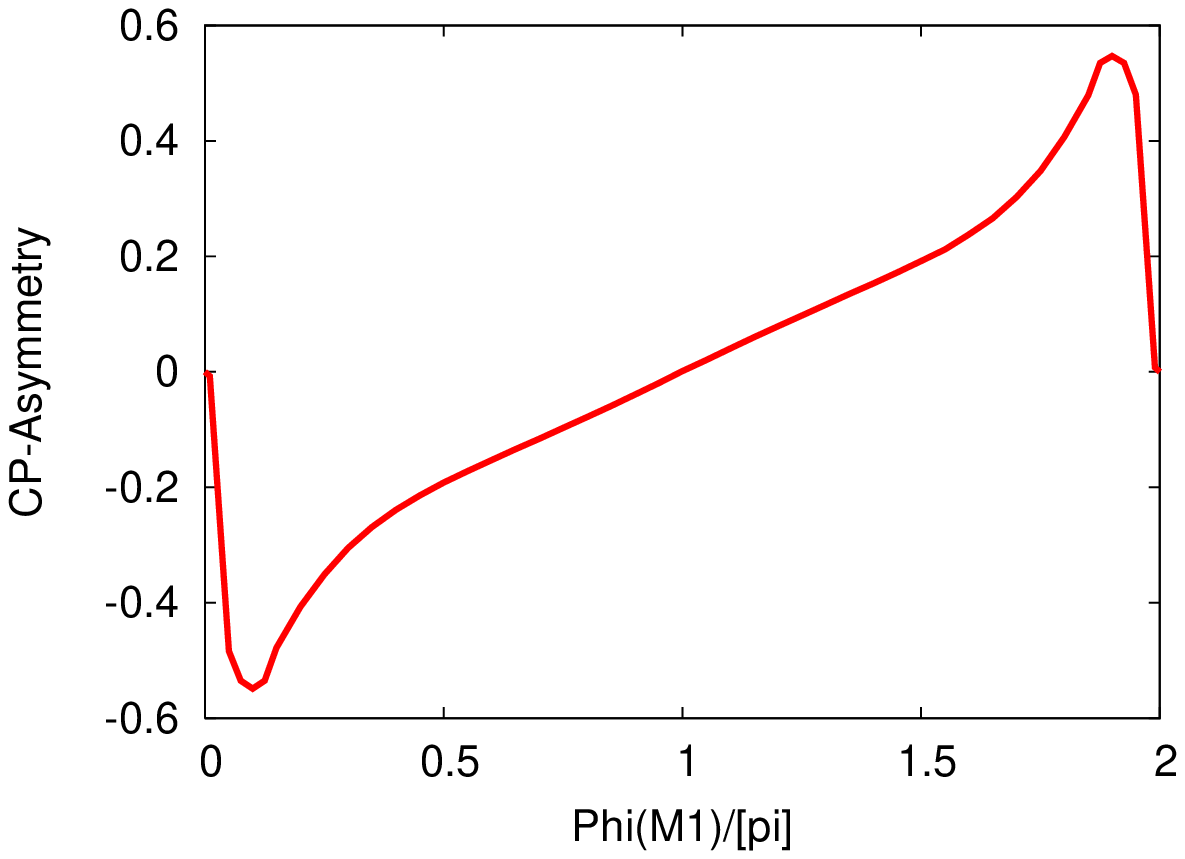}}
 \caption{
 Shapes of dilepton spectra in the cascade decays of new heavy particles at the LHC, for different sequences of spins (left)~\protect\cite{Webber}, and the sensitivity of a lepton asymmetry to a CP-violating phase in a
 supersymmetric model (right)~\protect\cite{EMM}. 
 }
\label{fig:LHC} 
\end{figure}

Spin effects in the decays of new particles produced at the LHC can be used to
analyze the underlying theory. For example, as already mentioned, spin plays a
central role in supersymmetry, which makes characteristic predictions for spin effects
in sparticle decay chains. A generic decay chain may include as many as four sparticles:
$D \to C \to B \to A$, each of which could in principle be vector (V), fermionic (F) or
scalar (S). In the case of squark decay, for example, the decay sequence could be
${\tilde q} \to \chi_2 [+q] \to {\tilde \ell} [+\ell] \to \chi [+\ell]$, with the spin signature SFSF. However, other
sequences and spin signatures could hold, e.g., in models with large extra dimensions.
There are many observables capable of distinguishing these spin signatures, such as
the shape of the $\ell \ell$ spectrum shown in the left panel of Fig.~\ref{fig:LHC}, 
the shape of the $q \ell$ spectrum and its angular
asymmetry as a function of its invariant mass~\cite{Webber}.

Such spin effects can also be used to look for CP-violating effects in sparticle decay.
Consider, for example, the case $gg \to {\tilde t}_1 {\tilde t}_1$ followed by ${\tilde t}_1 \to
\chi_2 +t$, followed by $\chi_2 \to \chi e^= e^-$. The lepton decay asymmetry is sensitive to
the CP-violating phase of the U(1) gaugino mass, as shown in the right panel of
Fig.~\ref{fig:LHC}~\cite{EMM}.

\subsection{Electron-positron collisions}

Polarized positrons (as well as electrons) would add significant value to the physics
programme of the ILC~\cite{polpos}. For example, in the direct channel, $RL$ and $LR$ collisions
select vector-boson exchanges, whereas $LL$ and $RR$ collisions select scalar
exchanges. Moreover, in many theories such as supersymmetry, the couplings of
particles that might be exchanged in the cross channel depend on the helicities of
the two colliding beams. the following are just a few examples of the gains for
studies of new physics that would be provided by having polarized positrons as well as electrons at the ILC.

$\bullet$ They would provide improved measurements of sparticle couplings, both by
providing new ways to suppress Standard Model backgrounds, e.g., from $W^+ W^-$, by
choosing suitable combinations of $e^\pm$ polarizations, and also by improving the
discriminating power for the quantum numbers of sparticles exchanged in the cross channel.

$\bullet$ They would provide improved sensitivities to new four-fermion contact interactions,
that could involve different combinations of $e^\pm$ helicities.

\begin{figure}
\centerline{\includegraphics[width=7.12cm]{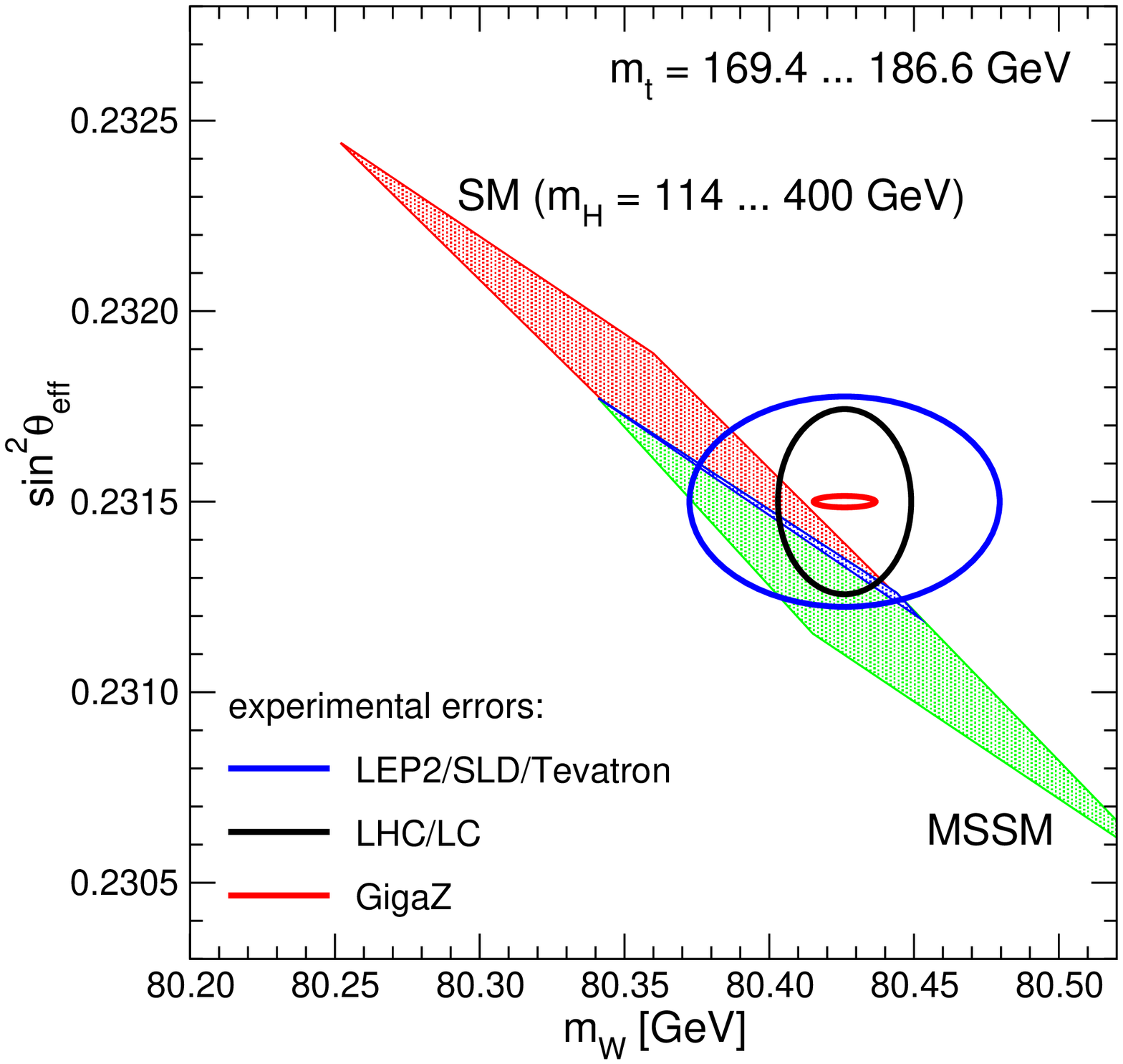}
\includegraphics[width=6.88cm]{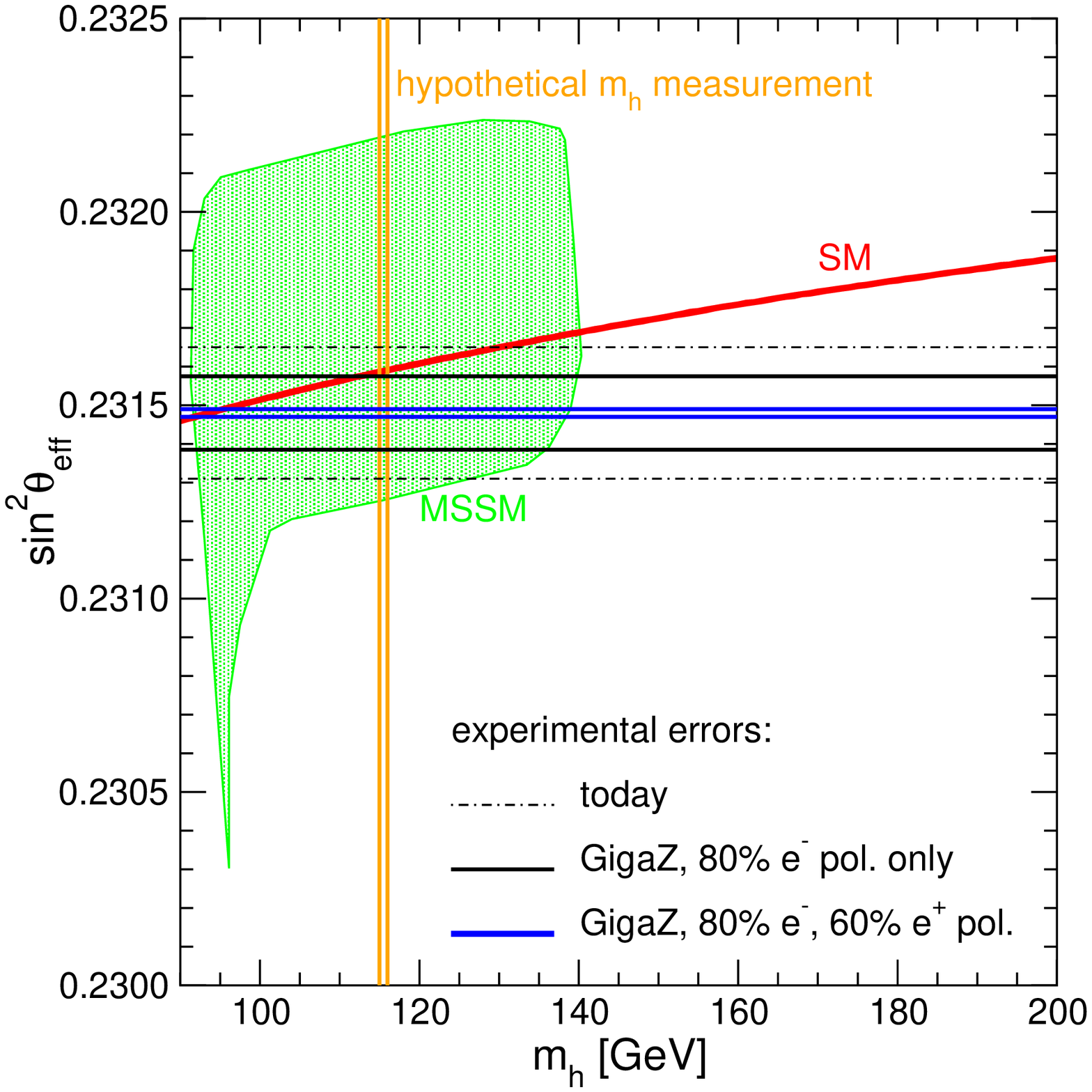}}
 \caption{
 The accuracies in $m_W$ and $\sin^2 \theta_W$ obtainable with the ILC Giga-$Z$ option,
 compared with the present and expected LHC/ILC accuracies (left), and the prospective
 accuracies of $\sin^2 \theta_W$ measurements with Giga-$Z$ with one or two
 polarized beams, compared with Standard Model and supersymmetric predictions as
 functions of the Higgs mass (right)~\protect\cite{HW}. 
 }
\label{fig:ILC} 
\end{figure}

$\bullet$ They would add value to the GigaZ programme of high statistics at the $Z$ peak~\cite{GigaZ}.
The measurements made possible by $e^\pm$ polarization would make possible improved
constraints on Standard Model parameters as well as enable quantum tests of possible
extensions of the Standard Model, such as supersymmetry, as seen in Fig.~\ref{fig:ILC}~\cite{HW}.

\section{The power of polarization}

The examples given in the previous sections illustrate some of the physics interest in
spin and polarization physics. It provides a unique tool for dissecting physics that we think 
we know, and often finds surprises, a prime example being the long-running puzzle of the
nucleon spin. Polarization can also be a delicate probe for new physics, by providing
new observables, suppressing backgrounds and enhancing signals. The probes of known
physics are complementary to the searches for new physics. Indeed, the
muon anomalous magnetic moment may already be providing a hint for new physics,
and electric dipole moments are excellent probes of CP violation. In
another example building on the
previous probes of the Standard Model at the $Z$ peak, 
future tests of its extensions using $e^\pm$ polarization at the ILC
will be very interesting. As another example,
our developing understanding of nucleon spin may eventually help us disentangle
the nature of dark matter. We must understand the Standard Model in order to probe beyond it,
and polarization is invaluable for both tasks.


\begin{thebibliography}{99}

\bibitem{Erler}
J.~Erler,
{\it Low energy tests of the standard model with spin degrees of freedom},
contribution to these proceedings,
  arXiv:hep-ph/0612030.
  %%CITATION = HEP-PH 0612030;%%
  
\bibitem{Chanowitz}
M.~S.~Chanowitz,
  %``Electroweak data and the Higgs boson mass: A case for new physics,''
  Phys.\ Rev.\ D {\bf 66}, 073002 (2002)
  [arXiv:hep-ph/0207123].
  %%CITATION = HEP-PH 0207123;%%
  I thank the author for conversations and the updated plot included here.

\bibitem{EWWG}
The LEP Collaborations and the LEP Electroweak Working Group,
{\it A combination of preliminary electroweak measurements and constraints  on
  the Standard Model},
arXiv:hep-ex/0612034.
  %%CITATION = HEP-EX 0612034;%%

\bibitem{GigaZ}
J.~Erler, S.~Heinemeyer, W.~Hollik, G.~Weiglein and P.~M.~Zerwas,
  %``Physics impact of GigaZ,''
  Phys.\ Lett.\ B {\bf 486}, 125 (2000)
  [arXiv:hep-ph/0005024].
  %%CITATION = HEP-PH 0005024;%%

\bibitem{NuTeV}
G.~P.~Zeller {\it et al.}  [NuTeV Collaboration],
  %``A precise determination of electroweak parameters in neutrino nucleon
  %scattering,''
  Phys.\ Rev.\ Lett.\  {\bf 88}, 091802 (2002)
  [Erratum-ibid.\  {\bf 90}, 239902 (2003)]
  [arXiv:hep-ex/0110059].
  %%CITATION = HEP-EX 0110059;%%
  
\bibitem{Moeller}
P.~L.~Anthony {\it et al.}  [SLAC E158 Collaboration],
  %``Precision measurement of the weak mixing angle in Moeller scattering,''
  Phys.\ Rev.\ Lett.\  {\bf 95}, 081601 (2005)
  [arXiv:hep-ex/0504049].
  %%CITATION = HEP-EX 0504049;%%
  
\bibitem{COMPASS}
V.~Y.~Alexakhin  [COMPASS Collaboration],
{\it The deuteron spin-dependent structure function g1(d) and its first
moment},
  arXiv:hep-ex/0609038.
  %%CITATION = HEP-EX 0609038;%%
  
\bibitem{HERMES}
HERMES Collaboration,
{\it Precise determination of the spin structure function g(1) of the proton,
deuteron and neutron},
  arXiv:hep-ex/0609039.
  %%CITATION = HEP-EX 0609039;%%

\bibitem{EK}
J.~R.~Ellis and M.~Karliner,
  %``Determination of alpha s and the nucleon spin decomposition using recent
  %polarized structure function data,''
  Phys.\ Lett.\ B {\bf 341}, 397 (1995)
  [arXiv:hep-ph/9407287].
  %%CITATION = HEP-PH 9407287;%%
  I thank Marek Karliner for conversations and the updated plot included here.

\bibitem{Mallot}
G.~K.~Mallot,
{\it Measurements of $\Delta G/G$}, contribution to these proceedings,
  arXiv:hep-ex/0612055.
  %%CITATION = HEP-EX 0612055;%%
  
\bibitem{Gabriela}
G.~A.~Navarro and R.~Sassot,
  %``Constraints on gluon polarization in the nucleon at NLO accuracy,''
  Phys.\ Rev.\ D {\bf 74}, 011502 (2006)
  [arXiv:hep-ph/0605266].
  %%CITATION = HEP-PH 0605266;%%

\bibitem{AAC}
M.~Hirai, S.~Kumano and N.~Saito,
{\it Constraint on $\Delta(g(x))$ from $\pi^0$ production at RHIC},
  arXiv:hep-ph/0612037.
  %%CITATION = HEP-PH 0612037;%%

\bibitem{COMPASS2}
  E.~S.~Ageev {\it et al.}  [COMPASS Collaboration],
  %``Gluon polarization in the nucleon from quasi-real photoproduction of
  %high-p(T) hadron pairs,''
  Phys.\ Lett.\ B  {\bf 633}, 25 (2006)
  [arXiv:hep-ex/0511028].
  %%CITATION = HEP-EX 0511028;%%

\bibitem{HERMES2}
P.~Liebig for the HERMES Collaboration, session 2A, these proceedings.

\bibitem{glueren}
A.~V.~Efremov and O.~V.~Teryaev,
  %``SPIN STRUCTURE OF THE NUCLEON AND TRIANGLE ANOMALY,''
JINR-E2-88-287;
G.~Altarelli and G.~G.~Ross,
  %``THE ANOMALOUS GLUON CONTRIBUTION TO POLARIZED LEPTOPRODUCTION,''
  Phys.\ Lett.\ B {\bf 212}, 391 (1988);
  %%CITATION = PHLTA,B212,391;%%
R.~D.~Carlitz, J.~C.~Collins and A.~H.~Mueller,
  %``THE ROLE OF THE AXIAL ANOMALY IN MEASURING SPIN DEPENDENT PARTON
  %DISTRIBUTIONS,''
  Phys.\ Lett.\ B {\bf 214}, 229 (1988).
  %%CITATION = PHLTA,B214,229;%%
  
\bibitem{JM}
R.~L.~Jaffe and A.~Manohar,
  %``THE g(1) PROBLEM: FACT AND FANTASY ON THE SPIN OF THE PROTON,''
  Nucl.\ Phys.\ B {\bf 337}, 509 (1990).
  %%CITATION = NUPHA,B337,509;%%
  
\bibitem{STAR}
M.~Sarsour,
{\it Recent results from the STAR spin program at RHIC},
  arXiv:hep-ex/0612065.
  %%CITATION = HEP-EX 0612065;%%

\bibitem{HERMs}
A.~Airapetian {\it et al.}  [HERMES Collaboration],
  %``Quark helicity distributions in the nucleon for up, down, and strange
  %quarks from semi-inclusive deep-inelastic scattering,''
  Phys.\ Rev.\ D {\bf 71}, 012003 (2005)
  [arXiv:hep-ex/0407032].
  %%CITATION = HEP-EX 0407032;%%
  
\bibitem{Aram}
A.~Kotzinian,
  %``LEPTO and polarized SIDIS,''
  Eur.\ Phys.\ J.\ C {\bf 44}, 211 (2005)
  [arXiv:hep-ph/0410093].
  %%CITATION = HEP-PH 0410093;%%
  
\bibitem{hierarchy}
L.~Maiani, {\it All You Need To Know About The Higgs Boson}, Proceedings
of the Gif-sur-Yvette Summer School On Particle Physics, 1979, pp.1-52; G.~'t~Hooft, in {\it
Recent developments in Gauge Theories}, Proceedings of the NATO Advanced Study
Institute, Carg{\`e}se, 1979, eds. G.~'t~Hooft et al. (Plenum Press, NY, 1980);
E.~Witten,
  %``Mass Hierarchies In Supersymmetric Theories,''
  Phys.\ Lett.\ B {\bf 105}, 267 (1981).
  %%CITATION = PHLTA,B105,267;%%

\bibitem{GUT}
J.~R.~Ellis, S.~Kelley and D.~V.~Nanopoulos,
  %``Precision Lep Data, Supersymmetric Guts And String Unification,''
  Phys.\ Lett.\ B {\bf 249}, 441(1990) and
  %%CITATION = PHLTA,B249,441;%%
%``Probing the desert using gauge coupling unification,''
  Phys.\ Lett.\ B {\bf 260}, 131 (1991);
  %%CITATION = PHLTA,B260,131;%%
U.~Amaldi, W.~de Boer and H.~Furstenau,
  %``Comparison of grand unified theories with electroweak and strong coupling
  %constants measured at LEP,''
  Phys.\ Lett.\ B {\bf 260}, 447 (1991);
  %%CITATION = PHLTA,B260,447;%%
  C.~Giunti, C.~W.~Kim and U.~W.~Lee,
  %``Running coupling constants and grand unification models,''
  Mod.\ Phys.\ Lett.\ A {\bf 6}, 1745 (1991);
  %%CITATION = MPLAE,A6,1745;%%
P.~Langacker and M.~x.~Luo,
  %``Implications of precision electroweak experiments for M(t), rho(0),
  %sin**2-Theta(W) and grand unification,''
  Phys.\ Rev.\ D {\bf 44}, 817 (1991).
  %%CITATION = PHRVA,D44,817;%%
  
\bibitem{EHNOS}
J.~R.~Ellis, J.~S.~Hagelin, D.~V.~Nanopoulos and M.~Srednicki,
  %``Search For Supersymmetry At The Anti-P P Collider,''
  Phys.\ Lett.\ B {\bf 127}, 233 (1983);
  %%CITATION = PHLTA,B127,233;%%
  J.~R.~Ellis, J.~S.~Hagelin, D.~V.~Nanopoulos, K.~A.~Olive and M.~Srednicki,
  %``Supersymmetric relics from the big bang,''
  Nucl.\ Phys.\ B {\bf 238} (1984) 453;
  %%CITATION = NUPHA,B238,453;%%
H.~Goldberg,
  %``Constraint on the photino mass from cosmology,''
  Phys.\ Rev.\ Lett.\  {\bf 50}, 1419 (1983).
  %%CITATION = PRLTA,50,1419;%%
  
\bibitem{g-2}
G.~W.~Bennett {\it et al.}  [Muon G-2 Collaboration],
  %``Final report of the muon E821 anomalous magnetic moment measurement at
  %BNL,''
  Phys.\ Rev.\ D {\bf 73}, 072003 (2006)
  [arXiv:hep-ex/0602035].
  %%CITATION = HEP-EX 0602035;%%

\bibitem{Davier}
M.~Davier, S.~Eidelman, A.~Hocker and Z.~Zhang,
  %``Updated estimate of the muon magnetic moment using revised results from  e+
  %e- annihilation,''
  Eur.\ Phys.\ J.\ C {\bf 31}, 503 (2003)
  [arXiv:hep-ph/0308213].
  %%CITATION = HEP-PH 0308213;%%

\bibitem{ICHEP}
D.~Wood, Plenary Talk at the 33rd International Conference on High-Energy Physics, Moscow, August 2006, {\tt http://ichep06.jinr.ru/session.asp?sid=1}.

\bibitem{susyg-2}
See, for example, J.~R.~Ellis, D.~V.~Nanopoulos and K.~A.~Olive,
  %``Combining the muon anomalous magnetic moment with other constraints on  the
  %CMSSM,''
  Phys.\ Lett.\ B {\bf 508}, 65 (2001)
  [arXiv:hep-ph/0102331].
  %%CITATION = HEP-PH 0102331;%%

\bibitem{Pospelov}
K.~A.~Olive, M.~Pospelov, A.~Ritz and Y.~Santoso,
  %``CP-odd phase correlations and electric dipole moments,''
  Phys.\ Rev.\ D {\bf 72}, 075001 (2005)
  [arXiv:hep-ph/0506106].
  %%CITATION = HEP-PH 0506106;%%

\bibitem{Bench}
M.~Battaglia {\it et al.},
  %``Proposed post-LEP benchmarks for supersymmetry,''
  Eur.\ Phys.\ J.\ C {\bf 22} (2001) 535
  [arXiv:hep-ph/0106204];
  %%CITATION = HEP-PH 0106204;%%
M.~Battaglia, A.~De Roeck, J.~R.~Ellis, F.~Gianotti, K.~A.~Olive and L.~Pape,
  %``Updated post-WMAP benchmarks for supersymmetry,''
  Eur.\ Phys.\ J.\ C {\bf 33} (2004) 273
  [arXiv:hep-ph/0306219].
  %%CITATION = HEP-PH 0306219;%%

\bibitem{Orlov}
Y.~F.~Orlov, W.~M.~Morse and Y.~K.~Semertzidis,
  %``Resonance method of electric-dipole-moment measurements in storage
  %rings,''
  Phys.\ Rev.\ Lett.\  {\bf 96}, 214802 (2006)
  [arXiv:hep-ex/0605022].
  %%CITATION = HEP-EX 0605022;%%

\bibitem{EFR}
J.~R.~Ellis, R.~A.~Flores and S.~Ritz,
  %``Implications for dark matter particles of searches for energetic solar
  %neutrinos,''
  Phys.\ Lett.\ B {\bf 198}, 393 (1987).
  %%CITATION = PHLTA,B198,393;%%

\bibitem{EOSS}
J.~R.~Ellis, K.~A.~Olive, Y.~Santoso and V.~C.~Spanos,
  %``Update on the direct detection of supersymmetric dark matter,''
  Phys.\ Rev.\ D {\bf 71}, 095007 (2005)
  [arXiv:hep-ph/0502001].
  %%CITATION = HEP-PH 0502001;%%
  
\bibitem{GMF}
R.~Gaitskell, V. Mandic and J. Filippini, \\
{\tt http://dendera.berkeley.edu/plotter/entryform.html}.

\bibitem{tWdecay}
D. Glenzinski, Plenary Talk at the 33rd International Conference on High-Energy Physics, Moscow, August 2006, {\tt http://ichep06.jinr.ru/session.asp?sid=1}.

\bibitem{Webber}
C.~Athanasiou, C.~G.~Lester, J.~M.~Smillie and B.~R.~Webber,
  %``Distinguishing spins in decay chains at the Large Hadron Collider,''
  JHEP {\bf 0608}, 055 (2006)
  [arXiv:hep-ph/0605286].
  %%CITATION = HEP-PH 0605286;%%

\bibitem{EMM}
J.~R.~Ellis, F.~Moortgat and G.~A.~Moortgat-Pick, in preparation.

\bibitem{polpos}
G.~A.~Moortgat-Pick {\it et al.},
  %``The role of polarized positrons and electrons in revealing fundamental
  %interactions at the linear collider,''
  arXiv:hep-ph/0507011.
  %%CITATION = HEP-PH 0507011;%%

\bibitem{HW}
S.~Heinemeyer and G.~Weiglein,
  %``Leading electroweak two loop corrections to precision observables in the
  %MSSM,''
  JHEP {\bf 0210}, 072 (2002)
  [arXiv:hep-ph/0209305].
  %%CITATION = HEP-PH 0209305;%%

\end{thebibliography}
\end{document}